# A Residual Network based Deep Learning Model for Detection of COVID-19 from Cough Sounds

*Annesya Banerjee, Achal Nilhani*

Department of Electronics and Telecommunication Engineering, Jadavpur University, India
{banerjee.annesya1999, achalnilhani010898}@gmail.com

## Abstract

The present work proposes a deep-learning-based approach for the classification of COVID-19 coughs from non-COVID-19 coughs and that can be used as a low-resource-based tool for early detection of the onset of such respiratory diseases. The proposed system uses the ResNet-50 architecture, a popularly known Convolutional Neural Network (CNN) for image recognition tasks, fed with the log-Mel spectrums of the audio data to discriminate between the two types of coughs. For the training and validation of the proposed deep learning model, this work utilizes the Track-1 dataset provided by the DiCOVA Challenge 2021 organizers. Additionally, to increase the number of COVID-positive samples and to enhance variability in the training data, it has also utilized a large open-source database of COVID-19 coughs collected by the EPFL CoughVid team. Our developed model has achieved an average validation AUC of 98.88%. Also, applying this model on the Blind Test Set released by the DiCOVA Challenge, the system has achieved a Test AUC of 75.91%, Test Specificity of 62.50%, and Test Sensitivity of 80.49%. Consequently, this submission has secured 16[th] position in the DiCOVA Challenge 2021 leader board.

## 1. Introduction

The sudden outbreak of the Corona Virus Diseases (COVID-19) in December 2019 has imposed an unprecedented and tremendous load on the healthcare system worldwide. Strength, scalability, and preparedness of the health-care industry has become the key factor in deciding the nations' ability to combat the pandemic. Extensive efforts are being made by researchers across the globe to apply technological solutions as an aid to the excessive pressure on the medical-care system. Emerging technologies including Artificial Intelligence (AI), Robotics, and Automation are being used for the production of affordable, agile testing solutions, contact tracing, implementing social distancing, development of lung-CT diagnosis tools [1], [2]. Different wearable technologies have shown promising applications in the domain of telehealth, patient monitoring, and disease detection [3] during the pandemic. Also, the different acoustics signals produced by the human body, that provide valuable information in detecting the diseases, have been analyzed. [4], [5] discuss in detail the use cases of acoustic signals (e.g., audio, speech, language) in combination with computer audition amidst the corona crisis. One such effort is the early diagnosis of the COVID-19 disease in the human body through cough symptom identification. As per the guidelines provided by World Health Organization (WHO) [6], coughs, particularly the dry and persistent ones, are among the most common symptoms in COVID-19 patients. Almost 67.7% of patients have been reported to develop cough symptoms, according to an early report [7] of WHO. Considering the fact that different diseases affect the respiratory system differently and thus produce cough signals of distinct acoustic patterns [8], several efforts have been made over the last few years to devise affordable diagnosis tools to identify a variety of respiratory diseases from analysis of cough sounds [9], [11], [12], [13], [14], [15], [16]. Since the outbreak of the coronavirus, such cough-based diagnosis efforts have been focused primarily to identify COVID-19 symptoms. The authors of [17] have developed an AI-engine implemented in a smart-phone app named AI4COVID-19 to analyze the cough sounds recorded by the user and identify if it is caused by the coronavirus disease. A similar approach is adopted in [18] to develop a respiratory sound (cough and breath) based automatic diagnosis tool for COVID-19 detection. The study has used large-scale crowd-sourced data collected through android and web applications to perform the analysis. The works in [19], [20], [21] also pivot around the deployment of a mobile-phone-based COVID-19 cough assessment system. The DiCOVA Challenge [10] organized by the Project Coswara group is a similar approach that aims at accelerating research in diagnosing COVID-19 using acoustics (mainly cough sounds, as given in Track 1). The proposed method in this paper addresses the task of Track-1, i.e., to discriminate COVID-19 coughs from Non-COVID coughs by using a deep learning model fed with log-Mel-spectrum features of the audio signal. Our main contribution to this ongoing research on audio-based detection of COVID-19 disease is that we have developed a classification mechanism that yields satisfactory results while being of sufficiently low complexity and suitable for implementation in low-resource devices. The rest of the paper is arranged as follows: Section 2 discusses the proposed algorithmic approach, section 3 describes the datasets being used, section 4 reports the experimental results and finally, section 5 concludes the work.

## 2. Algorithmic Approach

Figure 1 represents the overall workflow of the proposed method. Details of each of the blocks are provided in the following sub-sections.

### 2.1. Audio Preprocessing

To avoid any machinery noise (e.g., audio recording device ON/OFF sound, clicking sound), the input audio data is first preprocessed. This preprocessing step consists of two main tasks: 1) Identifying the portion of the audio containing cough sounds and 2) Centre-cropping the 'coughing' segment of the audio to tackle with variable length audio data present in the training/validation dataset. To accomplish task 1 of this step, we have utilized a Voice Activity Detector (VAD) [25], [26], considering the fact that all non-cough sounds in the audio data are noises or undesired artifacts. The VAD calculates the probability $P_r$ of presence of desired (cough) sound in each 100 ms non-overlapping window of the audio data. The earliest window having $P_r$ greater than a predefined threshold probability $P_{TH}$ is considered as the starting point of the preprocessed audio data. Audio segments before that window are completely discarded. In this case, setting $P_{TH}$ equal to 0.6 yielded satisfactory results. Furthermore, since the provided data contained mostly cough sounds, and no overlapping speech data, the current work has not focused on separating coughs from interfering speech sounds in real scenarios.

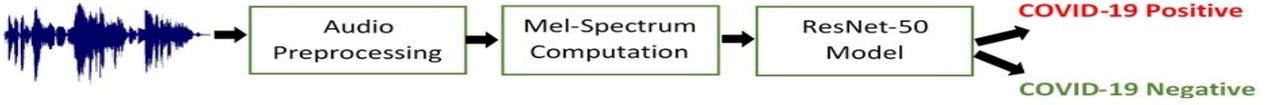

Figure 1: *Overall workflow diagram of the proposed method*

### 2.2. Mel-Spectrum Computation

Mel-scale features have been used widely in audio analysis tasks [22], from speech recognition, speaker recognition to musical genre classification, due to its perceptually motivated scales that resemble human-auditory perception. The Mel-scale frequency ($f_{Mel}$) is related to the normal frequency ($f_{Hz}$) using the relation in (1):

$$f_{Mel} = 2595 log_{10}(1 + \frac{f_{Hz}}{700}) \qquad (1)$$

so that the low frequencies (below 1kHz) are linearly spaced whereas high frequencies (above 1kHz) are logarithmically spaced. To obtain the Mel-spectrum, the Fourier transformed acoustic signal (X(k)) is passed through the Mel-filter bank having evenly spaced center frequencies of the filters and triangular filter shaper, in general. The Mel-spectrum $S_{Mel}(m)$ corresponding to the frequency domain signal X(k) is given by:

$$S_{Mel}(m) = \sum_{k=0}^{N-1}[|X(k)|^2 H_m(k)], \ 0 \leq m \leq M-1, \qquad (2)$$

where M is the total number of Mel-filters and $H_m(k)$ denotes the weight given to the $k^{th}$ energy bin to output the $m^{th}$ band. Figure 2 represents the Mel-spectrum of the COVID-19 positive and non-COVID-19 cough samples, where the spectrum is computed using M = 128 frequency bands over the frequency range of 32 Hz to 8 kHz with a sampling frequency of 16 kHz. As can be observed, there is a significant difference in the Mel-spectrums of the two different coughs, and therefore, using this feature, it is possible for a machine learning model to discriminate between these two inputs.

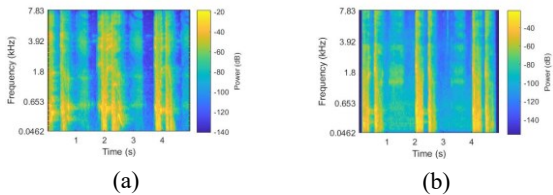

(a)                  (b)

Figure 2: *The Mel-spectrum of (a) COVID-19 positive and (b) non-COVID-19 coughs*

### 2.3. Deep Learning Model

Motivated by the success of the Residual Learning frameworks [23] in Image Recognition and Computer Vision tasks, we have used this deep Convolutional Neural Network (CNN) architecture for our purpose of audio classification. Analogous to the image identification tasks where the input to the model is an image, the input for this audio classification task is an image equivalent of the audio data. In other words, the Mel-spectrum itself works as a visual representation, similar to an image, for the time-frequency components of the input audio. Here, it should be noted that instead of directly feeding the Mel-Spectrum to the CNN, we use the log-Mel-Spectrum which successfully tackles the wide variation in audio spectrums, in general. Figure 3 represents the detailed architecture of the Residual Network of 50-layers depth (ResNet-50) along with the expected size of input and output data. For input with audio Mel-Spectrum or log-Mel-spectrum, input size should be $M \times K \times 1$, M being the number of frequency bands (e.g., 128) and K being the total number of time-frames for analysis (here, 155 for a 5s long audio segment; 1024 points window and 512 points overlap is considered). The network outputs the probabilities of the unknown audio segment being COVID-19 or Non-COVID-19.

### 3. Dataset

For the training and validation of the proposed model, we have primarily used the Track-1 dataset provided by the DiCOVA Challenge 2021 organizers. This dataset contains 1040 audio samples that are divided into 5-fold training-validation sets (overlapping) as per the challenge rule. Each audio is accompanied by the patient's gender (Male - m/Female – f), nationality (India – I/ Other – O), and COVID status (positive/negative) data. However, the number of COVID-19 positive samples is extremely low, compared to the Non-COVID ones in this dataset. To address this class imbalance problem, our team initially relied on Data Augmentation techniques, including pitch shifting, speedup, time-shifting, variation of volume, the addition of noise, etc. that helps in training the model over a variety of data representations. But the class imbalance being very large (only 75 COVID samples compared to 965 non-COVID samples), data augmentation was not sufficient to achieve a satisfactory performance of the deep learning model. Therefore, for the training of our system, we have also utilized the open-source dataset, collected and released by the EPFL CoughVid team [19]. This additional dataset from EPFL consists of a total of 20072 audio samples out of which 1010 samples are from patients who have been diagnosed as COVID-19 positive. Also, each audio sample in this dataset is accompanied by the probability of that audio sample being a cough. For our purpose, we have considered only those audio samples whose probability of being cough is more than 0.6. Furthermore, from this subset of audio samples, we have only used the confirmed COVID-positive samples. Using these selection rules, we have finally obtained 640 additional COVID samples from this EPFL dataset. These data along with the Track-1 Challenge dataset have been used for training the proposed log-Mel-spectrum+ResNet-50 model. For testing the model, the DiCOVA Blind Test Set with 233 audio samples has been used.

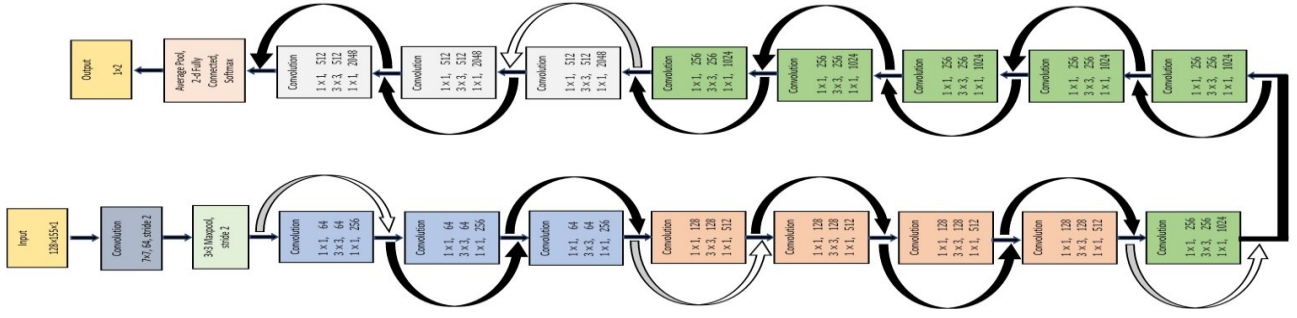

Figure 3: *The Residual Network – 50 Architecture with log-Mel-Spectrum as the input. Solid fill arrows ( ) denote direct addition (when input and output have same dimension). White fill arrows ( ) denote a convolution followed by addition operation (when input and output have different dimensions).*

# 4. Experimental Results

## 4.1. Evaluation Metrics

For evaluating the proposed model performance, the following metrics have been used:

$$Accuracy\ (ACC) = \frac{TP + TN}{TP + TN + FP + FN}$$
$$Sensitivity\ (SE) = \frac{TP}{TP + FN}$$
$$Specificity\ (SP) = \frac{TN}{TN + FP}$$
$$Precision\ (PR) = \frac{TP}{TP + FP}$$
$$F1 - Score\ (F1) = 2 \times \left(\frac{PR \times SE}{PR + SE}\right)$$

where, TP, TN, FP, FN denotes True Positive, True Negative, False Positive, and False Negative, respectively. Also, the Receiver Operating Characteristics (ROC) curve has been plotted to evaluate the Area Under ROC Curve (AUC) parameter. To adhere to the rules of the DiCOVA competition, the evaluation parameters have been calculated based on the mean confusion charts from the 5-fold cross-validation.

## 4.2. Network Training

The composite system of Figure 1 is trained using the combined dataset with a 90%-10% non-overlapping train-validation data split. Although pre-trained versions of the ResNet-50 are publicly available, we preferred training from scratch to adapt to the uniqueness of the present problem. Adam stochastic optimizer [24] is used to train the network because of its advantages including reduced memory requirement, computational efficiency, and little requirement of hyperparameter tuning. The initial learning rate of 0.0001, mini-batch size of 20, and a total of 25 epochs are set for training. For better training, the order of the training samples is shuffled randomly after every epoch. All the model training and evaluation tasks in this work have been executed in a system with Intel® Core ™ i7-10510U CPU, 16 GB RAM, and NVIDIA GeForce MX250 GPU.

## 4.3. Baseline Models

### 4.3.1. DiCOVA Baseline [10]

The DiCOVA baseline system, provided by the challenge organizers, extracts the Mel-frequency-cepstral-coefficients (MFCC) and its derivates, as well as double derivatives – all total 39-dimensional features. This feature set is used to train three different machine learning classifiers: Logistic Regression (LR), Multi-Layer Perceptron (MLP), and Random Forest (RF). For the training of both LR and MLP, $l_2$ penalty regularization is used. Additionally, MLP contains 25 hidden units followed by a tanh ( ) activation layer. On the other hand, during training the RF classifier, *gini* impurity criteria is used for the measurement of split quality. The best performance (average validation AUC of 68.54% and test AUC of 69.85%, as reported by the organizers) is obtained using the RF classifier and therefore, only this classifier will be used for further comparisons.

### 4.3.2. Brown et al. [18]

The authors propose a Logistic Regression Model (LRM) for the discrimination of cough symptoms from people who tested COVID-19 positive and people with no such 'COVID-19 positive' result. The model takes as input the temporal features like Duration, Tempo, Onset, Root Mean Square energy, Spectral features, and MFCC features along with its derivatives. In the case of spectral features, the statistical parameters (like mean, standard deviation, etc.) of the feature distribution are used in place of the exact feature values. The proposed model has been trained by a large-scale crowdsourced dataset collected by the authors. This work reports that the LRM has achieved a ROC-AUC of 0.75(0.19) with a precision of 0.82(0.20) and a recall of 0.54(0.23); the values in ( ) denote the standard deviations.

## 4.4. Results

Here we present the evaluation results obtained from our proposed model and also compare these results with the baseline model performances. It should be noted that the ground-truth labels for the blind test set have not yet been made available publicly. Thus, most of the evaluation performances presented in this section are calculated based on the 5-fold cross-validation dataset.

In the first experiment, we identify the relation between the number of frequency bands (M) used for the analysis of the audio and the deep learning model performance. The M values have been varied from 32 to 512 in powers of 2. Figure 4 represents the Confusion Matrices obtained from the ResNet50 model for five different values of M. As can be observed, the highest classification accuracy of 96% is obtained for M = 32 and the accuracy value continuously decreases as M is increased. The five different ROC curves, shown in figure 5, also depict that M = 32 yields the best performance whereas M = 512 produces the worst performance. The AUC values for M = 32, 64, 128, 256 and 512 are 98.88%, 98.21%, 94.36%, 93.59%, and 86.24%, respectively.

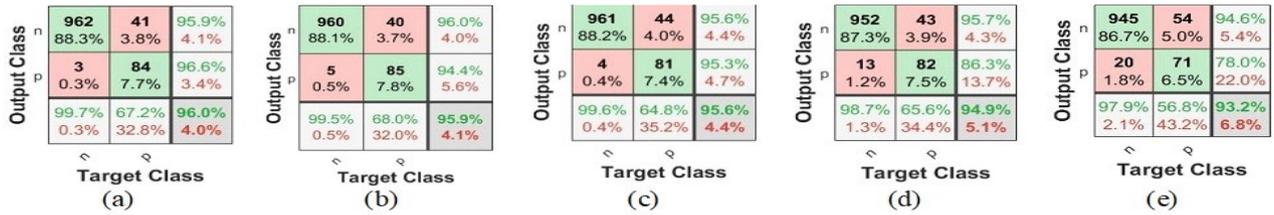

Figure 4: *Confusion matrix based on validation dataset for frequency bands (a) 32 (b) 64 (c) 128 (d) 256 and (e) 512*

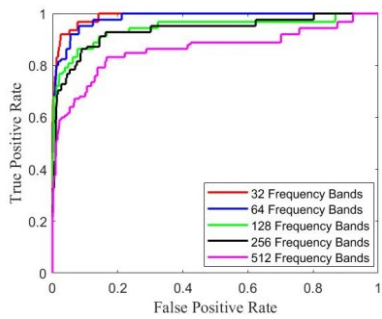

Figure 5: *ROC Curves for five different Frequency band-based ResNet-50 models*

A possible explanation for this degrading model performance with increasing M is as follows. Keeping the overall frequency range of the input audio constant, if M is increased then the dimension of each time-frequency bin reduces, and therefore, the audio feature information carried by each bin is also reduced. This, ultimately, results in poor classification performance of the deep learning model. However, making M very low (in the range of 4 to 16) is also not helpful because the size of analysis bins becomes so large that extracting important information becomes complicated. Thus, the optimum value of M lies in the range of 32 to 64. We take M = 32 as the best case and proceed further.

Table 1: *Comparison of different model performances*

| Ref # | Evaluation Metrics | | | | |
|---|---|---|---|---|---|
| | ACC | SE | SP | PR | F1 |
| 10 | 88.6 | **77.5** | 91.6 | 71.7 | 74.5 |
| 18 | 95.1 | 57.6 | **100** | **100** | 73.1 |
| Our Work | **96.0** | 67.2 | 99.7 | 96.6 | **79.3** |

In the next experiment, we compare the performance of the proposed model with the baseline systems. For this purpose, the baseline models mentioned in section 4.3. have been trained and validated using the same data as that of the proposed model. Whenever possible, the model parameters have been kept the same as that mentioned in the original papers [10], [18]. In other cases, standard values are assumed. Table 1 depicts the performance evaluation metrics yielded by these models; all values are in percentage (%) unless mentioned otherwise. As can be noted, the highest accuracy of 96% and F1 – score of 79.3% is obtained for the proposed model. Whereas, the model in [18] outperforms in terms of specificity and precision. However, none of the three systems achieve a very satisfactory performance in terms of sensitivity. This can be a major drawback for practical implementations and thus, needs to be improved in future works.

Also, from the submission results (using the blind test set) obtained from the organizer, it is observed that the proposed system achieves a Test AUC of 75.91%, Test Specificity of 62.50%, and Test Sensitivity of 80.49%. Consequently, our submitted model has secured 16th position in the DiCOVA Challenge 2021 leader board.

## 5. Conclusions

To conclude, the present work proposes a deep learning model based on the ResNet-50 architecture for the discrimination of COVID-19 and non-COVID-19 coughs. The model accepts 32 frequency band-based log-Mel-spectrums of the audio data as 'image'-like inputs and executes the two-class classification task. The developed model has achieved an average validation AUC of 98.88%. Also, applying this model on the blind test set released by the DiCOVA challenge, the system has achieved a Test AUC of 75.91%, Test Specificity of 62.50%, and Test Sensitivity of 80.49%. Further improvements in the system will lead to better sensitivity of the classification model and enable the possibility of real-world implementation.

## 6. Acknowledgments

The authors would like to thank the DiCOVA Challenge 2021 organizers for making the crowdsourced dataset of COVID-19 coughs available for use by the scientific community and letting the authors contribute to this research progress for early detection of COVID-19 affected people.

## 7. References

[1] Tulin Ozturk, Muhammed Talo, Eylul Azra Yildirim, Ulas Baran Baloglu, Ozal Yildirim, and U Rajendra Acharya. Automated detection of covid-19 cases using deep neural networks with x-ray images. Computers in Biology and Medicine, page 103792, 2020.

[2] Halgurd S Maghdid, Kayhan Zrar Ghafoor, Ali Safaa Sadiq, Kevin Curran, and Khaled Rabie. A novel ai-enabled framework to diagnose coronavirus covid 19 using smartphone embedded sensors: Design study. arXiv preprint arXiv:2003.07434, 2020.

[3] Xiao-Rong Ding, David Clifton, JI Nan, Nigel Hamilton Lovell, Paolo Bonato, Wei Chen, Xinge Yu, Zhong Xue, Ting Xiang, Xi Long, et al. Wearable sensing and telehealth technology with potential applications in the coronavirus pandemic. IEEE Reviews in Biomedical Engineering, 2020.

[4] Björn W Schuller, Dagmar M Schuller, Kun Qian, Juan Liu, Huaiyuan Zheng, and Xiao Li. Covid-19 and computer audition: An overview on what speech & sound analysis could contribute in the sars-cov-2 corona crisis. arXiv preprint arXiv:2003.11117, 2020.

[5] Gauri Deshpande and Björn Schuller. An overview on audio, signal, speech, & language processing for covid-19. arXiv preprint arXiv:2005.08579, 2020.

[6] World Health Organization et al. Coronavirus disease 2019 (covid-19): situation report, 70. 2020.

[7] World Health Organization, World Health Organization, et al. Report of the who-china joint mission on coronavirus disease 2019 (covid-19), 2020.

[8] William Thorpe, Miranda Kurver, Greg King, and Cheryl Salome. Acoustic analysis of cough. In The Seventh Australian and New Zealand Intelligent Information Systems Conference, 2001, pages 391–394. IEEE, 2001.

[9] Hwan Ing Hee, BT Balamurali, Arivazhagan Karunakaran, Dorien Herremans, Onn Hoe Teoh, Khai Pin Lee, Sung Shin Teng, Simon Lui, and Jer Ming Chen. Development of machine learning for asthmatic and healthy voluntary cough sounds: A proof of concept study. Applied Sciences, 9(14):2833, 2019.

[10] Muguli, Ananya, Lancelot Pinto, Neeraj Sharma, Prashant Krishnan, Prasanta Kumar Ghosh, Rohit Kumar, Shreyas Ramoji et al. DiCOVA Challenge: Dataset, task, and baseline system for COVID-19 diagnosis using acoustics. arXiv preprint arXiv:2103.09148, 2021.

[11] Renard Xaviero Adhi Pramono, Syed Anas Imtiaz, and Esther RodriguezVillegas. A cough-based algorithm for automatic diagnosis of pertussis. PloS one, 11(9): e0162128, 2016.

[12] Roneel V Sharan, Udantha R Abeyratne, Vinayak R Swarnkar, and Paul Porter. Automatic croup diagnosis using cough sound recognition. IEEE Transactions on Biomedical Engineering, 66(2):485–495, 2018.

[13] Anthony Windmon, Mona Minakshi, Pratool Bharti, Sriram Chellappan, Marcia Johansson, Bradlee A Jenkins, and Ponrathi R Athilingam. Tussiswatch: A smart-phone system to identify cough episodes as early symptoms of chronic obstructive pulmonary disease and congestive heart failure. IEEE journal of biomedical and health informatics, 23(4):1566–1573, 2018.

[14] Charles Bales, Charles John, Hasan Farooq, Usama Masood, Muhammad Nabeel, and Ali Imran. Can machine learning be used to recognize and diagnose coughs? arXiv preprint arXiv:2004.01495, 2020.

[15] Yusuf Amrulloh, Udantha Abeyratne, Vinayak Swarnkar, and Rina Triasih. Cough sound analysis for pneumonia and asthma classification in pediatric population. In 2015 6th International Conference on Intelligent Systems, Modelling and Simulation, pages 127–131. IEEE, 2015.

[16] Keegan Kosasih, Udantha R Abeyratne, Vinayak Swarnkar, and Rina Triasih. Wavelet augmented cough analysis for rapid childhood pneumonia diagnosis. IEEE Transactions on Biomedical Engineering, 62(4):1185–1194, 2014.

[17] Ali Imran, Iryna Posokhova, Haneya N Qureshi, Usama Masood, Sajid Riaz, Kamran Ali, Charles N John, Iftikhar Hussain, and Muhammad Nabeel. Ai4covid-19: AI enabled preliminary diagnosis for covid-19 from cough samples via an app. Informatics in Medicine Unlocked, page 100378, 2020.

[18] Chloë Brown, Jagmohan Chauhan, Andreas Grammenos, Jing Han, Apinan Hasthanasombat, Dimitris Spathis, Tong Xia, Pietro Cicuta, and Cecilia Mascolo. Exploring automatic diagnosis of covid-19 from crowdsourced respiratory sound data. arXiv preprint arXiv:2006.05919, 2020.

[19] CoughVid. https://coughvid.epfl.ch/about/, note= Accessed: March 23, 2021.

[20] COVID Voice Detector. https://cvd.lti.cmu.edu/, note= Accessed: March 23, 2021.

[21] Neeraj Sharma, Prashant Krishnan, Rohit Kumar, Shreyas Ramoji, Srikanth Raj Chetupalli, Prasanta Kumar Ghosh, Sriram Ganapathy, et al. Coswara–a database of breathing, cough, and voice sounds for covid-19 diagnosis. arXiv preprint arXiv:2005.10548, 2020.

[22] Lawrence Rabiner and Ronald Schafer. Theory and applications of digital speech processing. Prentice Hall Press, 2010.

[23] He, Kaiming, Xiangyu Zhang, Shaoqing Ren, and Jian Sun. Deep residual learning for image recognition. In Proceedings of the IEEE conference on computer vision and pattern recognition, pages 770-778, 2016.

[24] Diederik P Kingma and Jimmy Ba. Adam: A method for stochastic optimization. arXiv preprint arXiv:1412.6980, 2014.

[25] Sohn, Jongseo., Nam Soo Kim, and Wonyong Sung. A Statistical Model-Based Voice Activity Detection. Signal Processing Letters IEEE, 6(1), 1999.

[26] Martin, Rainer. Noise power spectral density estimation based on optimal smoothing and minimum statistics. IEEE Transactions on speech and audio processing, 9(5): 504-512, 2001.